\documentclass[usenatbib,referee]{mn2e}
\usepackage{graphicx}

\def\deg{\ifmmode^\circ\else$^\circ$\fi}

\title[Post-Outburst Phase of McNeil's Nebula (V1647 Orionis)]
{Post-Outburst Phase of McNeil's Nebula \mbox{(V1647 Orionis)}}

\author[D.K. Ojha et al.]{D.K. Ojha,$^{1}$\thanks{E-mail: ojha@tifr.res.in} 
S.K. Ghosh,$^{1}$ A. Tej,$^{1}$ R.P. Verma$^{1}$ and S. Vig$^{1}$\\
$^{1}$Tata Institute of Fundamental Research, Mumbai (Bombay) - 400 005, India
\newauthor
G.C. Anupama,$^{2}$ D.K. Sahu,$^{2}$ P. Parihar,$^{2}$ B.C. Bhatt,$^{2}$ 
T.P. Prabhu,$^{2}$ 
\newauthor
G. Maheswar$^{2}$ and H.C. Bhatt$^{2}$\\
$^{2}$Indian Institute of Astrophysics, Koramangala, Bangalore - 560 034, India
\newauthor
B.G. Anandarao$^{3}$ and V. Venkataraman$^{3}$\\
$^{3}$Physical Research Laboratory, Ahmedabad - 390 009, India
}

\begin{document}

\date{}
\pagerange{\pageref{firstpage}--\pageref{lastpage}} \pubyear{2005}

\maketitle

\label{firstpage}

\begin{abstract}

We present a detailed study of the post-outburst phase 
of McNeil's nebula (V1647 Orionis) using 
optical $B,V,R,I$ and near-infrared (NIR) $J,H,K$ photometric and 
low resolution optical spectroscopic observations. 
The observations were carried out with the HFOSC, NIRCAM, 
TIRCAM and \mbox{NICMOS} cameras on the 2m HCT and 1.2m PRL telescopes during the period 2004 
February -- 2005 December. The optical and NIR observations show a general decline 
in brightness of the exciting source of McNeil's nebula (V1647 Ori). Our 
recent optical images show that V1647 Ori has faded by more than 3
magnitudes since February 2004.   
McNeil's nebula itself has also faded considerably. 
The optical/NIR photometric data also show a significant variation in the 
magnitudes 
($\Delta V$ = 0.78 mag, $\Delta R$ = 0.44 mag, $\Delta I$ = 0.21 mag, 
$\Delta J$ = 0.24 mag and $\Delta H$ = 0.20 mag) of V1647 Ori
within a period of one month, which is possibly undergoing
a phase similar to eruptive variables, like EXors or FUors. 
The optical spectra show a few features such as 
strong $H\alpha$ emission with blue-shifted absorption  
and the Ca II IR triplet (8498\AA, 8542\AA~and 8662\AA) in emission.
As compared to the period just after outburst, there is a decrease in
the depth and extent of the blue-shifted absorption component,
indicating a weakening in the powerful stellar wind.
The presence of the \mbox{Ca II} IR triplet in emission 
confirms that V1647 Ori is a pre-main-sequence star.
The long-term, post-outburst photometric observations 
of V1647 Ori suggest an EXor, rather than an FUor event.
An optical/IR comparison of the region surrounding McNeil's nebula 
shows that the optical 
nebula is more widely and predominantly extended to the north, 
whereas the IR nebula is relatively confined (diameter $\sim$ 60\arcsec), 
but definitely extended, to the south, too. The large 
colour gradient from north to south and the sudden absence of an optical nebula 
to the south is suggestive of a large scale disk-like structure (or envelope) 
surrounding the central source that hides the southern nebula. 

\end{abstract}


\section{Introduction}

On the night of 2004 January 23, Jay McNeil discovered a new reflection
nebula in the L1630 cloud in Orion (McNeil 2004). The nebula, which now 
bears his name, is recognized as a reflection nebula surrounding
a young stellar object (YSO). It appears
to be a faint optical counterpart of the infrared (IR) source
catalogued as IRAS 05436-0007 that had gone into outburst, producing a 
large reflection nebulosity (Reipurth and Aspin 2004; \'Abrah\'am et al. 2004;
Brice\~no et al. 2004; Vacca, Cushing and Simon 2004; Andrews, Rothberg and Simon
2004; Walter et al. 2004; Kun et al. 2004; Ojha et al. 2005). 
Later the McNeil's object was designated as V1647 Orionis (Samus 2004). 
Reipurth and Aspin (2004) reported that NIR images taken with
the Gemini 8m telescope show that the object has 
brightened by about 3 magnitudes in $JHK$, relative to the 1998 2MASS
measurements. Gemini spectra reveal strong features of CO and hydrogen 
Br$\gamma$ in emission in the IR. In the optical, $H\alpha$ appears in 
emission with a P Cygni profile (Reipurth and Aspin 2004).   

McGehee et al. (2004) described the change in optical and NIR
colours of V1647 Ori from its ``quiescent'' state to its outburst. They concluded 
that the colour changes associated with brightening suggest an EXor like outburst 
rather than a simple dust clearing event. Kun et al. (2004) analyzed optical 
and NIR measurements of V1647 Ori and suggested that the outburst 
was due to an intrinsic increase in the 
luminosity of the central object (rather than a decrease in extinction), and 
identified the object as an FUor, rather than an EXor. Vacca, Cushing and Simon 
(2004) presented NIR spectroscopy of the object and suggested that 
V1647 Ori is a heavily embedded low-mass Class I protostar, surrounded by 
a disk, whose brightening is due to a recent accretion event.
\'Abrah\'am et al. (2004) described IR and sub-mm properties of 
V1647 Ori
and concluded that the object has a disk of about 0.5 M$_{\odot}$, a 
pre-outburst bolometric luminosity of 5.6 L$_{\odot}$, and is probably a 
class II protostar of age about 0.4 Myr. Andrews, Rothberg and Simon (2004)
presented the measurements of V1647 Ori 
in the mid-IR and sub-mm and
concluded that the 12 $\mu$m flux of this source has increased by a factor of 
25 after the outburst, whereas the sub-mm continuum remains at its 
pre-outburst level. The infrared slope of the spectral energy distribution (SED)
characterizes the illuminating source as a flat-spectrum protostar, in both 
its active and quiescent states. Muzerolle et al. (2005) report on {\it Spitzer}
NIR photometry of V1647 Ori in its early post-outburst phase (March 2004). They 
found that the brightness of the star increased by roughly the same factor of 
15--20 across the optical and NIR. Their results indicated that the outburst may
be intermediate between FUor and EXor-type events.
Aspin and Reipurth (2005) reported on observations made at the Gemini telescope
and showed that V1647 Ori has faded by approximately one magnitude in $r'$-band
since January 8, 2005. K\'osp\'al et al. (2005) reported the most recent $VRI$ 
measurements of V1647 Ori obtained between 2005 October 4 and November 19 which
show that V1647 Ori has started a very rapid fading ($\sim$ 1 mag/month). They
concluded that V1647 Ori is a unique (somewhat intermediate) object among FUors and 
EXors. 

The McNeil's source at the apex of the nebula was also observed to brighten 
dramatically in X-ray wavelength.
The X-ray data from {\it Chandra} gives strong evidence that the probable 
cause of the outburst is the sudden infall of gas onto the surface of the 
star from an orbiting disk of gas (Kastner et al. 2004). Grosso et al. (2005) 
describe X-ray spectra of V1647 Ori taken 
with {\it XMM-Newton} on 2004 April 4. 
They find variations in both the brightness and hardness of the source. 
There is a very exciting hint of periodicity in the X-ray brightness, which might 
provide information on the physical nature of the X-ray luminosity. 
Vig et al. (2004, 2005) presented post-outburst radio continuum observations 
of the region around McNeil's nebula and report no radio continuum emission 
from the nebula. They explored various scenarios to explain the 
radio upper limit of the nebula and suggested that the homogeneous 
HII region scenario is consistent with the radio, $H\alpha$ as well as X-ray 
measurements. 

In this paper we present the optical and NIR morphologies of McNeil's nebula
and report on the optical/NIR photometric and optical spectroscopic monitoring of 
V1647 Ori beginning with
the epoch of the recent outburst. Our results provide variability 
measurements of V1647 Ori for a duration of about 21 months and allow us to 
check if the central star has changed significantly since its emergence in  
early 2004. 
Our optical/NIR observations also 
show significant variations in the source brightness within periods 
of one week and one month. In \S 2 we present the details of observations and 
data reduction procedures, \S 3 deals with the results and discussion 
concerning the short and long time variability of V1647 Ori that illuminates 
McNeil's nebula, and describes the optical/IR nebula in detail. We also
discuss spectroscopic results in this section. We then 
summarize our conclusions in \S 4.  

\section{Observations and Data Reduction}

\subsection{Optical Photometry}

Imaging observations of McNeil's nebula region at optical wavelengths 
(standard Bessel $BVRIH\alpha$ filters) were obtained on 19 nights during  
2004 February -- 2005 November
with the 2m 
Himalayan Chandra Telescope (HCT) situated at the Indian Astronomical 
Observatory (IAO), Hanle, India. The Himalaya 
Faint Object Spectrograph Camera (HFOSC) instrument equipped with a 
SITe 2K$\times$4K pixel CCD was used. The central 2K$\times$2K region, 
used for imaging, corresponds to a field of view (FOV) of 
$\sim$ 10\arcmin$\times$10\arcmin, 
with a pixel scale of 0.296\arcsec. Further details on the 
telescope and the instrument can be found at
\\ 
\mbox{http://www.iiap.res.in/iao/hfosc.html}. 
The observations were carried out under good photometric sky conditions.
The average seeing (FWHM) in all the 
bands was $\sim$ 1.6\arcsec~during the observations. Bias and flat frames were 
obtained at the beginning and end of each observing night. Atmospheric 
extinction and 
transformation coefficients were obtained from the observations of photometric 
standard stars around 98-556, 98-618, 98-961, PG 1525-071 and PG 0231+051 
regions (Landolt 1992). The details of the observations are given in Table 1.
Landolt standard regions were not observed on five nights, hence 
the magnitudes of V1647 Ori were obtained by using a reference star as 
a secondary standard in the field (see Table 1).
During five nights, observations were carried out for the target in $R$ filter
only and a reference star (see \S 3) in the field was used to obtain the 
differential magnitudes. 

Data reduction was done using the NOAO's IRAF\footnote{IRAF is distributed 
by the National Optical Astronomy Observatories, which are operated by the 
Association of Universities for Research in Astronomy, Inc., under contract 
to the National Science 
Foundation.} package tasks. Flat-fielding with a median flat frame was 
applied. Identification and photometry of point sources were performed by 
using the DAOFIND 
and DAOPHOT packages in IRAF, respectively. Because of nebulosity around 
V1647 Ori,
photometry was obtained using the point-spread function (PSF) algorithm 
ALLSTAR in the DAOPHOT package (Stetson 1987). The residuals to the photometric 
solution are $\le$0.05 mag. The absolute optical photometry of V1647 Ori is 
presented in Table 2.

\subsection{NIR Photometry}

The $JHK$ NIR observations were carried out during 2004 February 12, 
October 18, November 19, 2005 November 28, 29 and 30 
with the HCT NIR camera (NIRCAM), equipped with 512$\times$512 HgCdTe array of 
18 $\mu$m pixel size. For our observations, the NIRCAM was used in the 
{\it Camera-B} mode which 
has an FOV of 3.6\arcmin$\times$3.6\arcmin. Further details on NIRCAM 
can be found at \mbox{http://www.iiap.res.in/iao/nir.html}.
We obtained several dithered exposures of the target 
centred at ($\alpha$, $\delta$)$_{2000}$ = ($05^h46^m14^s.0$,
\mbox{-00$^{\circ}05^{\arcmin}40^{\arcsec}.1$)} in each of the NIR bands.
Typical integration times per frame were 20s, 15s and 5s in the $J$, $H$ and 
$K$-bands, respectively. The images were co-added to obtain the final image in each 
band. 
We also obtained several dithered sky frames close to the target position in each 
NIR band.
Photometric calibration was done from the
observations made on the same nights on the standard stars around AS13 region
(Hunt et al. 1998) at 
air masses close to that of the target 
observations. We used an aperture radius of 18 pixels ($\sim$ 7\arcsec) for 
photometry of V1647 Ori. 
The local sky was evaluated in an annulus 
with an inner radius of 128 pixels and a width of 12 pixels.

The $JHK$ NIR observations were also carried out on 2004 October 12, using TIFR 
Near-Infrared Camera (TIRCAM) at the f/13 Cassegrain focus of the 
Mount Abu 1.2m telescope belonging to PRL (India). 
TIRCAM is 
based on the SBRC InSb Focal Plane Array (58$\times$62 pixels) sensitive 
between 1--5 $\mu$m. The details of the TIRCAM system is described in Ojha et al. (2002). 
Individual object frames were of 2s integration in $J$, $H$ and $K$-band. 
Photometric calibration was obtained by observing the standard 
star HR1852 from the list of UKIRT's bright standards\footnote{http://www.jach.hawaii.edu/UKIRT/astronomy/calib/phot$_{-}$cal/bright$_{-}$stds.html}. 

We also made NIR $JHK'$ photometric 
observations by using PRL NICMOS camera on Mount Abu 1.2m telescope.
At the core of the NICMOS camera is a 256x256 HgCdTe array with
pixel scale of 1.0\arcsec ~and an FOV of 4\arcmin ~x 4\arcmin. 
Photometric observations were made in $JHK'$-bands on 10 occasions, 
namely 2004 March 20 \& 21, April 11, October 16 \& 17, December 21 \&
22, 2005 November 25 \& 26 and December 3 under photometric conditions. 
Single frame integration times
were 10/20 sec for $J$, 5/10 sec for $H$ and 2 sec for $K'$-band. Overall 
integration times were 300--360s for $J$, 200--300s for $H$ and 200--320s 
for $K'$-bands. Sufficient number of dithered sky frames were also taken along 
with dark frames. Photometric calibration was done by using the standard stars 
around AS13 region (Hunt et al. 1998) and HD 40335. 

The data reduction was done using IRAF software tasks. All the NIR images went 
through the standard pipeline reduction procedures like dark/sky-subtraction and 
flat-fielding. Photometric magnitudes were extracted by using the IRAF 
DAOPHOT/PHOT and APPHOT/PHOT tasks. 
NIR photometry of V1647 Ori
obtained on 17 nights 
(2004 February -- 2005 December) with NIRCAM, TIRCAM and NICMOS cameras 
is listed in Table 3 along with other published results in the literature. 

\subsection{Optical Spectroscopy}

Spectroscopic observations of V1647 Ori were carried out with 
the HFOSC instrument at the 2m HCT on 10 occasions beginning 
2004 February 22. 
The log of observations is given in Table 4. To avoid the contamination from 
nebular emission around McNeil's nebula, the slit was positioned on V1647 Ori 
in EW direction. The spectra were obtained at resolutions of $\sim$ 15\AA 
~and $\sim$ 7\AA ~in the 
wavelength range 5000--9100\AA. These resolutions are sufficient to resolve the 
$H\alpha$ line detected in the spectrum of V1647 Ori. The spectra were reduced 
using standard IRAF 
routines. The one-dimensional spectra were extracted from the bias-subtracted and 
flat-field corrected images using the optimal extraction method. Wavelength 
calibration of the spectra was done using the  
FeAr and FeNe lamp 
sources. Subsequently, the spectra were corrected for instrumental 
response using spectrophotometric standards (Hiltner 600, Feige 34 
and G191B2B) observed on the same night and brought to a flux scale.  
For the nights of 2005 September 8 and 28, the multiple flux 
calibrated spectra were combined and scaled to a weighted mean.

\section{Results and Discussion}

\subsection{Optical and IR Morphology of McNeil's Nebula}

A colour composite image was constructed from the HFOSC $V$, $H\alpha$ and 
$I$-band images ($V$ represented in blue, $H\alpha$ in green and $I$ in red) 
obtained on 2004 February 12 and is shown in figure 1. The location of V1647 Ori
is marked at the centre of the image. The well known Herbig-Haro (HH) objects,
such as the HH 24 complex, are seen in green colour because of the $H\alpha$
emission. The diffuse $H\alpha$ emission due to HH 23 is highlighted by the
box at top centre. The star associated with McNeil's nebula was proposed as the
exciting source of HH 23 (Eisloffel and Mundt 1997; Kastner et al. 2004).
A bright knot HH 22 is marked by an arrow, which is seen as a ``smoky cloud'' 
covering part of the new object.

The enlarged view of McNeil's nebula region is shown in figure 2. Figure 2a
shows the central region of nebula in optical wavelengths.  
The right panel (figure 2b) shows the $JHK$ 
composite colour image ($J$: blue, $H$: green and $K$: red) of McNeil's nebula 
observed
on 2004 October 18. The 
cometary infrared nebula is clearly seen extending 
towards north from the NIR source (V1647 Ori) and has a cavity structure with 
possibly two rims extending toward north-east and north-west (see also 
Ojha et al. 2005). The north-east rim is 
brighter and sharp, while the north-west rim is diffuse and is not seen clearly. 
The north-east rim can be traced out to
$\sim$ 30\arcsec~or $\sim$ 12000 AU at a surface brightness of 
$K$ $\sim$ 18.0 mag arcsec$^{-2}$ (where we assume that the distance to 
McNeil's nebula is 400 pc; 
Anthony-Twarog 1982). Such kind of morphology can probably be seen when the eruption 
has cleared out some of the dust and gas in the envelope surrounding the young star, 
allowing light to escape and to illuminate a cone-shaped cavity carved out by 
previous eruptions (Reipurth and Aspin 2004). 

In figure 3a, we show a contour diagram of the region around V1647 Ori, as seen 
in our $K$-band image obtained on 2004 October 18. The star is surrounded by a 
compact reflection nebulosity and there is a southward continuation of the 
extended nebula in NIR, which is demonstrated 
in the PSF-subtracted $K$-band image of figure 3b. This was confirmed by 
comparing the PSF of the NIR central source (V1647 Ori) with the PSF of the 
star located about 30\arcsec~south-west of V1647 Ori (figure 3b). We find 
that the compact IR nebula that encloses all of the nebular structure, 
including the two rims, has a diameter of about 60\arcsec. 

An optical/IR comparison of the region surrounding McNeil's nebula (see figures 
2 and 4) shows that the position of the bright infrared source is coincident with 
that of an optical source at the apex of the nebula that is illuminating the 
fan-shaped cloud of gas, or nebula (Reipurth and Aspin 2004). The optical nebula is 
more widely and predominantly extended to the north, whereas the IR nebula is
relatively confined, but definitely extended to the south, too.
The greenest portion of the optical nebula is not seen in IR including the 
knot HH 22 to the north-east and the green patch to the north (see figures 1, 
2 and 4). There is a clear 
colour gradient from north to south ($H-K$ colour varies from 0.53 to 1.33 from 
north to south), as well as from inside to outside of the cavity.
The gradient is very steep near the central star, and the southern extension of 
the nebula is seen only at IR. This large colour gradient and the absence of 
optical nebula to the south is suggestive of a large scale disk-like structure (or 
envelope) surrounding the central source that hides the southern nebula in
optical. This is 
often the case for YSOs associated with a cometary nebula (e.g., Tamura et al. 
1991).

\subsection{Optical/IR Variability of V1647 Ori}

Optical and NIR photometric results of V1647 Ori obtained on 27
nights (2004 February -- 2005 December) are listed in Tables 2 and 3. 

The optical photometry shows a general decline of brightness in V1647 Ori 
both in the beginning and later part of our observations
with an intermediate brightening phase. In our dataset V1647 Ori is 
brightest on 2004 October 18 (see Table 2). McNeil's nebula itself has now
faded considerably 
(see Table 2). A significant variation in the source brightness 
($\Delta V$ = 3.43 mag, $\Delta R$ = 3.36 mag and $\Delta I$ = 3.14 mag)
can be seen during a period of about 21 months (2004 February -- 2005 November; 
Table 2). 
We can also see a considerable variation in the source brightness 
($\Delta V$ = 0.78 mag, $\Delta R$ = 0.44 mag and $\Delta I$ = 0.21 mag) 
during a period of one month (2004 October -- November; Table 2). 
The difference in magnitudes is much larger than the photometric errors
($\le$ 0.05 mag). 
For comparison, the $BVRI$ magnitudes of the reference \mbox{star A} 
(($\alpha$, $\delta$)$_{2000}$ = $05^h46^m22^s.4$,
-00$^{\circ}$03$^{\arcmin}$37$^{\arcsec}$; marked in figure 1)
are listed in Table 5 along with other results from the literature, 
which match quite well with our measurements. 
As can be seen in Table 5, we do not see any significant variation 
($>$ 0.05 mag, which is the estimated photometric error)
in the magnitudes of the reference \mbox{star A}.
Hence, 
the brightness variation seen in 
V1647 Ori during a period of one month (2004 October -- November) is real. It 
is also seen that the variability is larger at shorter wavelengths 
compared to the longer ones (Table 2). 

The differential light curve (2004 February -- 2005 November) of V1647 Ori in 
$R$-band is shown in figure 5. Semkov (2004) reported the $VRI$ photometric 
observations of V1647 Ori and found that the source varied with amplitude of 
0.5 mag during the period 2004 August 18 -- October 3. 
In figure 5 we have also plotted the $R$-band light curve from Semkov (2004). 
The optical light curve shows a variable nature of V1647 Ori.
After a significant increase in the brightness around 2004 October 18, 
a general decline in the source brightness can be seen in recent observations.
Over the one and half year that we followed the 
star (2004 February -- 2005 November), the brightness has decreased by more 
than 3 magnitudes in $R$-band. The shapes of the light curves are similar 
in other optical bands. 

A comparison of the $JHK$ magnitudes (see Table 3) of V1647 Ori shows that the 
source has varied over the time of observation between 2MASS (1998 October) 
and HCT/PRL (2005 November/December).  
We also see a significant variation in the source 
brightness ($\Delta J$ = 0.13 mag, $\Delta H$ = 0.26 mag, and 
$\Delta K$ = 0.14 mag) during a period of one week (2004 October 12 -- 18).
A similar kind of weekly variation in NIR magnitudes of V1647 Ori 
during the early phase of outburst (2004 February 23 -- 29) was reported by 
Ojha et al. (2004, 2005). The short time variability in optical and NIR 
wavelengths in V1647 Ori was also noticed by Walter et al. (2004). They 
found a general decline in brightness of about 0.3--0.4 magnitudes during 
87 days (2004 February 11 -- May 7). 

Figure 6 shows a $J-H/H-K$ colour-colour (CC) diagram. We notice changes of the 
infrared colours of V1647 Ori from before eruption (1998) to after the 
eruption (2004 - 2005). A general movement
along a reddening vector can be seen 
during 1998 -- 2005.
The source had moved toward the tip of the locus of classical T Tauri stars 
(Meyer et al. 1997) during 2004. This suggests that V1647 Ori has NIR colour 
similar to a dereddened T Tauri star. The $J-H$ and $H-K$ colours indicate 
that the circumstellar matter of $A_V$ $\sim$ 8 mag was probably cleared in 
the eruption, in less than six years since the 2MASS observations. 
This scenario is consistent with the present IR morphology of the McNeil's 
nebula region (see \S 3.1). There are colour changes between the most recent 
measurements (2005 November) and those close to peak brightness (2004). The
source has now become redder and fainter and moved back along a reddening 
vector in the CC diagram.     
 
The brightness increase in all the optical and NIR bands near the
outburst peak (early 2004) tells that the accretion disk luminosity
that depends directly on the accretion rate
must have gone up. As the enhanced acceretion has decreased over a period
of about two years the disk luminosity too has decreased. So it may be
predominantly the stellar plus the original disk excess as seen 
in the infrared. The position in the
CC diagram therefore is a combination of extinction and thermal dust
emission. As the object is embedded in dust and gas there will also be
some reflection effects.

\subsection{Spectral Features}

The optical spectra of V1647 Ori obtained with HCT are shown in figure 7.  
The spectra show a very red continuum with a number of emission features. 
The $H\alpha$ and Ca II IR triplet lines are 
in emission. The presence of the Ca II IR triplet in emission
confirms that V1647 Ori is a pre-main-sequence star.
We also identified a few lines of the Paschen series of hydrogen 
at low signal-to-noise ratio level
(e.g. Pa 23, Pa 20, Pa 19, Pa 18, Pa 17, Pa 12, Pa 11 and Pa 10)
in our deep spectra of 2005 September 8 and 28 
in the wavelength range 8340 -- 9020\AA. In our spectra,
the OI $\lambda7773$ line is seen in absorption.
The detection of OI $\lambda$7773 
line indicates that the optical light probably originates in the low-gravity, 
hot photosphere of the inner disk (Walter et al. 2004).
Since our spectra are not 
corrected for telluric features, we see the absorptions
at about 7600\AA~ and 8230\AA~due to atmospheric O2-A and H$_2$Oz bands,
respectively. In the spectrum taken on 2005 September 8, we 
detected the Mg I line at 8807\AA
~and Fe I at 8824\AA ~rather marginally.
Table 6 shows the equivalent width (EW) measurements of $H\alpha$,
\mbox{OI $\lambda$7773} and Ca II IR triplet from HCT spectra. We estimate the
measurement uncertainties in the EW to be of the order of $\sim$ 2\%.
The EW of 
the Ca II triplet are nearly same 
which is in contrast with the results of Walter et al. (2004), where
the ratio of the triplet EWs was shown to be 2:2:1. Interestingly 
Walter et al had used a resolution similar to ours. It must be noted 
here that in our data the Ca II triplet lines have not been de-blended for 
the HI Pa 13, 15 and 16 lines. Since the other Paschen lines are not seen 
as prominently as the towering Ca II triplet, we do not expect significant 
contamination in Ca II triplet by Pa 13, 15 and 16 lines.

Due to the somewhat low/medium resolution, we can not resolve the actual 
FWHMs and peak heights of the components of the triplet which would have 
probably thrown some light on the possible optical depth variations within the 
Ca II emission regions. In general the Ca II triplet arises from optically 
thick regions of electron densities of about 10$^{9}$--10$^{10}$ cm$^{-3}$, 
with temperatures between about 4000--10000 K (Ferland \& Persson 1989). 
Following the arguments of Hamann \& Persson (1992) for T Tauri stars, we
infer from the nearly equal EWs of Ca II triplet lines that the
collisional decay is more dominant process than radiative decay. This in turn
implies that the product of optical depth and electron density 
needs to be more than
or equal to about 10$^{13}$ cm$^{-3}$ (see Hamann \& Persson 1992). Coupling 
this argument with the absence of the [Ca II] lines at $\sim$ 7300\AA~
in the HCT spectra, we may conclude that the density of the Ca II region is
very large and is optically very thick.

On one occasion, namely 2005 September 8, the HCT spectra show clearly the 
Ca II forbidden doublet at 8912/8927\AA, while the other forbidden doublet 
at 7291/7323\AA~ is still absent.
These forbidden lines arise in optically thin regions in comparison with the 
Ca II permitted lines, otherwise they will be de-excited by collisions.
The doublet 8912/8927\AA ~is about 8.4 eV from the ground state while the 
doublet 7291/7323\AA ~is only 1.7 eV (e.g., Lang 1978). This is probably 
indicative of the high temperature of the chromosphere of the T Tauri type 
star that we are monitoring. This region must also have a lower density 
than the triplet region in order not to get de-excited by collisions.
Hamann \& Persson (1992) noticed that [Ca II] lines occur rarely in 
T Tauri stars and in some cases where they did, they showed blue-shifted 
profiles. Since the HCT spectrum is rather spiky in the region near the 
doublet, it is difficult to say if the line peak is Doppler shifted.

The HCT spectra show prominent broad $H\alpha$ line on all occasions with 
varying EW (see figure 8). The line seems to arise from regions having large 
scale turbulent velocities in comparison to the narrow lines of Ca II. The 
FWHMs of the $H\alpha$ emission lines (after correcting for instrumental 
broadening by using quadrature formula) show variations ranging
from 300 to 600 km/s. 
Such large widths are quite common in T Tauri
stars (e.g., Hamann \& Persson 1992). There appears to be a positive correlation 
between the $H\alpha$ and the Ca II EWs indicating that the two emitting regions
are responsive to each other (Hamann \& Persson 1992). We do
not see this correlation in \mbox{OI $\lambda$7773} line (in absorption) which
occurs in rather cooler regions in comparison.

The strong $H\alpha$ is in emission, 
with a P Cygni absorption feature seen in the early phase of outburst 
(2004 February 22 and 23; see figure 8), first reported by 
Reipurth and Aspin (2004). 
Similar P Cygni absorption features are also seen in the 
Ca II IR triplet in early outburst spectra of V1647 Ori.  
The pronounced 
P Cygni profile seen in $H\alpha$ is likely to be formed in a strong wind 
that has sufficient optical depth to produce the deep blue-shifted absorption 
trough. The well-defined blue edge of the absorption trough indicates 
a wind velocity of 420 -- 460 km/s (figure 8; see also Maheswar and Bhatt 2004).
This implies significant mass-loss in a powerful wind.
We notice a significant decrease in the
depth and the extent of this strong absorption component over the period of
our observations (figure 8),
as also reported
by Aspin and Reipurth (2005). This suggests weakening of the powerful wind.


Figure 9 shws the time variation of the $H\alpha$ EW over a period of about 
19 months, which includes data from Walter et al (2004). The data from 
Walter et al (2004) indicate an oscillatory or 
periodic variation of the $H\alpha$ EW. We also probably see such a variation 
in our dataset. 
Power spectrum analysis indicates the period of oscillations
to be $\sim$ 40 days.
This variation may probably be caused by temporal variations in the
wind/accretion flow or by the rotational period of V1647 Ori.  
However, poor time sampling prevents us from making definitive conclusions
about the periodic variation in $H\alpha$ EW. 


From Table 6 we notice that the strength of the H$\alpha$ emission line
has increased, whereas the 
OI $\lambda$7773 line in absorption 
appears quite strong in the early phase of outburst 
(2004 February -- November) and has 
become weaker (in fact it is not seen clearly on 2005 September 28). 
Compared to the early outburst stages,
Ca II IR triplet lines also show significant strengthening. 
Hamann and Persson (1992) have suggested that the
OI $\lambda$7773 absorptions are entirely due to warm gas in the envelope
and the presence of this line suggests either a rotating disk or a 
highly turbulent envelope. If it were an outburst then turbulence
in the envelope gas is expected initially which in turn gives stronger
OI $\lambda$7773 line. But as the time progresses, the turbulence may
die down and the line becomes weaker. Hence variability in the 
OI $\lambda$7773 line indeed indicates an outburst activity in the source.


  

\subsection{Possible EXor Event ?}

Two classes of erupting low-mass YSOs have emerged: FUors, which display outbursts
of $\sim$ 4 magnitudes and more that last several decades and EXors, which show 
smaller outbursts ($\Delta m$ $\sim$ 2--3 mag) that last from a few months to a 
few years and may occur repeatedly (Herbig 1977; Bell et al. 1995; 
Hartmann 1998).
V1647 Ori had brightened by about 3 magnitudes in $JHK$, relative to
the 1998 2MASS measurements and is probably returning to its pre-outburst state. 
It is therefore possible that we had witnessed an EX Orionis 
kind of behavior, since the source has about the correct amplitude 
(Reipurth and Aspin 2004; Ojha et al. 2005). 
The time elapsed since the outburst of V1647 Ori is more 
than two years now.
During this period our optical data show a significant variation in the 
source brightness (see Table 2); the latest optical images show that
V1647 Ori has faded by more than 3 magnitude since February 2004.  
We also notice a significant weekly and monthly variations in the brightness
of V1647 Ori over the period of our observations
(2004 February -- 2005 November/December; see Tables 2 and 3). 
Similar kind of variations are also seen from the
light curves of Walter et al. (2004). This seems to 
suggest that we are probably witnessing a sporadic or episodic
outburst (e.g., Herbig et al. 2001), which may indicate an EXor kind
of eruption of V1647 Ori. However, further photometric monitoring of the 
object is required to settle the class of eruptive variables to which 
V1647 Ori belongs. 

V1647 Ori has now started a very rapid fading. The object may
indeed be returning to its pre-outburst characteristics
though in the NIR it is still 1.3 -- 1.4 mag brighter than the 2MASS
(pre-outburst) brightness. At both optical and NIR, the object was
bluer and brighter near the outburst peak brightness and is now redder and
fainter. Such a behaviour would be expected if during the outburst a
disk, hotter than the stellar photosphere, dominated the luminosity
signifying high accretion rate. With the decline in the accretion, the
luminosity is now dominated by the star (cool late type) and a less
pronounced disk still emits some excess IR. 
This also shows that the outburst is essentially caused by accretion
enhancement.
  
 

\section{Conclusions}

In this paper we have presented long-term, post-outburst photometric
and spectroscopic monitoring of V1647 Ori. 
Following conclusions are drawn from our study:

1) The optical data show that V1647 Ori has faded by more than
3 magnitude since February 2004. We also see significant variations in the
brightness of V1647 Ori within a period of one week
($\Delta m$ $\sim$ 0.3 mag) and one month ($\Delta m$ $\sim$ 0.8 mag). This
shows the eruptive behaviour of McNeil's nebula.

2) McNeil's nebula has a cavity structure with possibly two rims as seen in NIR 
wavelengths. This kind of morphology might have been produced by the eruption 
that has cleared out a part of the envelope surrounding the young star.  
This 
scenario is supported by the $J-H$ and $H-K$ colours (near the outburst
peak brightness) of V1647 Ori which indicate that the circumstellar matter of 
$A_V$ $\sim$ 8 mag was probably cleared in the eruption.

3) An optical/IR comparison of McNeil's nebula shows that the optical nebula 
is more widely and predominantly extended to the north, while the IR nebula 
is relatively confined, but definitely extended, to the south, too. This 
colour gradient of the nebula is most likely due to the presence of a large 
scale disk-like structure (or envelope) around the young star.


4) The optical spectra show H$\alpha$ line strongly in emission and 
display a pronounced P Cygni profile, with an absorption trough reaching 
a velocity of $\sim$ 460 km/s during the early phase of outburst. 
The depth and the extent of the strong absorption component have 
decreased significantly over the period of our observations,
suggesting that the powerful wind is weakening.
Variation in $H\alpha$ line strength on the time scale of $\sim$ 1 day 
is also seen. 

5) The presence of strong emission lines, relatively fast fading
rate and the low accretion rate (10$^{-5}$ M$_{\odot}$/yr; Muzerolle
et al. 2005) of V1647 Ori suggest
probably the evidence for a new EXor event.
 
\section{Acknowledgements}
The observations reported in this paper were obtained using the 2m Himalayan 
Chandra Telescope at IAO Hanle, 
the high altitude station of the Indian Institute of Astrophysics, Bangalore. 
We thank the staff at IAO and at the remote control station at CREST, 
Hosakote for assistance during the observations. It is a pleasure to thank 
the staff at Mount Abu IR observatory for their support during the 
observations. We especially thank to the referee Dr. William Sherry for the
critical reading of our manuscript.

\clearpage

\begin{table}
\caption{Observing log of optical photometric observations}
\begin{tabular}{cccccc}
\\
\hline
\hline
Date (UT)   &    JD     & FWHM$^\dag$ & $^\pm$Filter(s) & Exposure time (in secs) & Notes\\
\hline
2004 Feb 12 & 2453048.1 & 1.2\arcsec & $R,I$     & 480, 240                 & 1\\
2004 Feb 13 & 2453049.1 & 2.3\arcsec & $R$       & 90                       & 2\\
2004 Feb 14 & 2453050.2 & 1.5\arcsec & $V,R,I$   & 200, 270, 90, 900        & ...\\
2004 Feb 22 & 2453057.9 & 1.4\arcsec & $V,R,I$   & 300, 150, 150            & ...\\
2004 Feb 23 & 2453058.9 & 1.7\arcsec & $B,R$     & 1200, 150 & 1\\
2004 Oct 18 & 2453297.4 & 2.3\arcsec & $B,V,R,I$ & 300, 300, 300, 300, 600  & ...\\
2004 Nov 17 & 2453327.4 & 1.4\arcsec & $B,V,R,I$ & 600, 900, 540, 360, 900  & ...\\
2004 Dec 7  & 2453347.3 & 1.8\arcsec & $B,V,R,I$ & 600, 720, 600, 480, 1200 & ...\\
2005 Jan 8  & 2453379.1 & 1.5\arcsec & $R$       & 120                      & 2\\
2005 Jan 9  & 2453380.1 & 2.1\arcsec & $R$       & 60                       & 2\\
2005 Mar 28 & 2453458.1 & 1.7\arcsec & $R$       & 900                      & 2\\
2005 Sep 26 & 2453639.6 & 1.5\arcsec & $V,R$     & 1200, 600                & ...\\
2005 Sep 29 & 2453643.5 & 1.4\arcsec & $B,V,R$   & 900, 450, 450            & ...\\
2005 Nov 9  & 2453684.3 & 1.3\arcsec & $V,R,I$   & 300, 440, 420            & 1\\
2005 Nov 20 & 2453695.2 & 2.0\arcsec & $V,R,I$   & 600, 450, 450            & 1\\
2005 Nov 21 & 2453696.4 & 1.3\arcsec & $V,R,I$   & 420, 420, 420            & 1\\
2005 Nov 28 & 2453703.2 & 1.5\arcsec & $V,R,I$   & 600, 1800, 720           & ...\\
2005 Nov 29 & 2453704.2 & 1.7\arcsec & $R$       & 1200                     & 2\\
2005 Nov 30 & 2453705.2 & 1.6\arcsec & $V,R,I$   & 600, 1200, 720           & ...\\
\hline
\end{tabular}

$^\dag$Measured average FWHM. This is a measure of the seeing.

$^\pm$Observations were made in Bessel $B,V,R,I$ filters.

$^1$Magnitudes of V1647 Ori were obtained by using a reference star as a 
secondary standard in the field.

$^2$Only differential photometry was done.
\end{table}


\begin{table}
\caption{Optical $BVRI$ photometry of V1647 Ori and McNeil's nebula at HCT}
\begin{tabular}{cccccccccc}
\\
\hline
\hline
Date (UT)  & \multicolumn{4}{|c|}{V1647 Ori$^\dag$} & & \multicolumn{4}{|c|}{McNeil's nebula$^\S$}\\\cline{2-5}\cline{7-10} 
\\
 &   $B$ &  $V$        &     $R$       &      $I$      &  &  $B$ & $V$ & $R$ & $I$\\
\hline
$^\pm$2004 Feb 12 & ... &  ...          & 16.91$\pm$.02 & 14.82$\pm$.02 & & ... &  ...          & 13.18$\pm$.02 & 12.13$\pm$.02 \\
2004 Feb 14 & ... & 18.68$\pm$.04 & 16.83$\pm$.02 & 14.86$\pm$.02  & & ... & 13.98$\pm$.04 & 13.19$\pm$.02 & 12.19$\pm$.02 \\ 
2004 Feb 22 & ...           & 18.85$\pm$.01 & 16.92$\pm$.02 & 14.81$\pm$.01 & & 14.54$\pm$.01 & 13.62$\pm$.01 & 12.98$\pm$.02 & 11.94$\pm$.01 \\
$^\pm$2004 Feb 23 & 20.44$\pm$.01 & ...           & 16.74$\pm$.01 & ...       &    & 14.44$\pm$.01 & ...           & 13.00$\pm$.01 & ...           \\
$^\ast$2004 Oct 18 & ... & 18.02$\pm$.05 & 16.51$\pm$.05 & 14.76$\pm$.05 &  & 14.77$\pm$.05 & 13.87$\pm$.05 & 13.16$\pm$.05 & 12.05$\pm$.05 \\
2004 Nov 17 & 20.64$\pm$.04 & 18.80$\pm$.02 & 16.95$\pm$.01 & 14.97$\pm$.02 & & 14.85$\pm$.04 & 13.87$\pm$.02 & 13.19$\pm$.01 & 12.09$\pm$.02 \\
2004 Dec 7  & 20.72$\pm$.05 & 19.08$\pm$.01 & 17.15$\pm$.02 & 15.13$\pm$.02 & & 14.82$\pm$.05 & 13.90$\pm$.01 & 13.26$\pm$.02 & 12.05$\pm$.02 \\
$^\ast$2005 Sep 26 &    ...        & 19.82$\pm$.02 & 17.94$\pm$.02 &  ...       &   &      ...        & 14.92$\pm$.02 & 14.12$\pm$.02 &  ...          \\
$^\ast$2005 Sep 29 &    ...        & 19.82$\pm$.01 & 18.09$\pm$.01 &  ...       &   &  15.88$\pm$.01 & 15.01$\pm$.01 & 14.11$\pm$.01 &  ...          \\
$^\pm$2005 Nov 9       &     ...         & 21.45$\pm$.05  & 19.46$\pm$.04  & 17.34$\pm$.05 & &  ... & 16.02$\pm$.05 & 14.94$\pm$.04 & 13.57$\pm$.05 \\  
$^\pm$2005 Nov 20      &     ...         & $>$ 21.45$^{\ast\ast}$  & 19.76$\pm$.04  & 17.61$\pm$.02 & &  ... &  16.66$\pm$.05       & 15.24$\pm$.04 & 13.72$\pm$.02 \\
$^\pm$2005 Nov 21      &     ...         & $>$ 21.45$^{\ast\ast}$  & 19.83$\pm$.02  & 17.56$\pm$.02 & &  ... & 16.96$\pm$.04        & 15.19$\pm$.02 & 13.58$\pm$.02\\
2005 Nov 28            &     ...         & 22.11$\pm$.05           &
20.02$\pm$.03  & 17.93$\pm$.05 & &  ... & 16.90$\pm$.05 &
15.14$\pm$.03 & 13.60$\pm$.05\\
2005 Nov 30 & ... & $>$ 22.11$^{\ast\ast}$ & 20.10$\pm$.05 & 17.95$\pm$.02 & & ... & 16.62$\pm$.05 & 15.18$\pm$.05 & 13.78$\pm$.02 \\
\hline
\end{tabular}

$^\pm$Magnitudes are derived using star A (see text) as a secondary calibrator.

$^\dag$PSF magnitudes.

$^\S$Magnitudes in 80\arcsec~aperture centred on V1647 Ori.

$^\ast$Too faint to be detected in $B$-band.

$^{\ast\ast}$Source is not detected in $V$-band with exposure time of 600s.

\end{table}


\begin{table}
\caption{NIR $JHK$ photometry of V1647 Ori}
\begin{tabular}{cccccc}
\\
\hline
\hline
Date (UT)   &     $J$         &     $H$         &  $K$   & FWHM & Telescope  \\
\hline
1998 Oct 7  & 14.74$\pm$.03 & 12.16$\pm$.03 & 10.27$\pm$.02 &  ...       & 2MASS \\
2004 Feb 3  & 11.1$\pm$.1   &  9.0$\pm$.1   &  7.4$\pm$.1   & 0.5\arcsec & Gemini$^\dag$\\
2004 Feb 11 & 10.79$\pm$.01 & 8.83$\pm$.01  & 7.72$\pm$.01  &  ...       & USNO$^\S$\\
2004 Feb 12 & 10.68$\pm$.04 & 8.78$\pm$.04  & ...           & 1.2\arcsec & HCT\\
2004 Feb 23 & 10.74$\pm$.05 & 8.93$\pm$.01  & 7.40$\pm$.02  & 1.6\arcsec & IRSF$^\ast$\\
2004 Feb 25 & 10.79$\pm$.05 & 8.95$\pm$.01  & 7.47$\pm$.01  & 0.9\arcsec & IRSF$^\ast$\\
2004 Feb 29 & 10.90$\pm$.06 & 9.11$\pm$.01  & 7.62$\pm$.02  & 1.4\arcsec & IRSF$^\ast$\\
2004 Mar 20 & 10.85$\pm$.07 & 8.83$\pm$.02  & 7.71$\pm$.05  & 1.7\arcsec & PRL/NICMOS\\
2004 Mar 21 & 10.76$\pm$.02 & 8.91$\pm$.01  & 7.59$\pm$.03  & 1.8\arcsec & PRL/NICMOS\\
2004 Apr 11 & 11.09$\pm$.03 & 9.01$\pm$.02  & 7.49$\pm$.03  & 2.5\arcsec & PRL/NICMOS\\
2004 Apr 12 & 10.94$\pm$.02 & 9.06$\pm$.02  & 7.59$\pm$.02  & ...        & USNO$^\S$\\
2004 Oct 12 & 10.76$\pm$.06 & 8.87$\pm$.06  & 7.34$\pm$.06  & 2.1\arcsec & TIRCAM\\
2004 Oct 16 & 11.15$\pm$.01 & 9.24$\pm$.02  & 7.89$\pm$.01  & 1.7\arcsec & PRL/NICMOS\\
2004 Oct 17 & 11.18$\pm$.02 & 9.23$\pm$.01  & 7.87$\pm$.02  & 1.7\arcsec & PRL/NICMOS\\
2004 Oct 18 & 10.89$\pm$.04 & 9.13$\pm$.04  & 7.48$\pm$.04  & 2.3\arcsec & HCT\\
2004 Nov 19 & 11.13$\pm$.04 & 9.33$\pm$.04  & $<$ 8.03      & 1.0\arcsec & HCT$^\pm$\\
2004 Dec 21 & 11.04$\pm$.03 & 9.20$\pm$.01  & 7.83$\pm$.03  & 2.3\arcsec & PRL/NICMOS\\
2004 Dec 22 & 11.03$\pm$.02 & 8.98$\pm$.01  & 7.76$\pm$.03  & 2.1\arcsec & PRL/NICMOS\\
2005 Nov 25 & ...           & 10.02$\pm$.07 & ...           & 2.5\arcsec & PRL/NICMOS\\
2005 Nov 26 & 13.28$\pm$.15 & 10.69$\pm$.09 & ...           & 2.5\arcsec & PRL/NICMOS\\
2005 Nov 28 & ...           & 10.80$\pm$.02 & 8.95$\pm$.03  & 1.5\arcsec & HCT\\
2005 Nov 29 & 13.05$\pm$.05 & 10.70$\pm$.05 & 8.86$\pm$.04  & 1.7\arcsec & HCT\\
2005 Nov 30 & 13.06$\pm$.02 & 10.70$\pm$.03 & 8.93$\pm$.02  & 1.6\arcsec & HCT\\
2005 Dec 3  & 13.45$\pm$.01 & 10.87$\pm$.01 & ...           & 1.9\arcsec & PRL/NICMOS\\
\hline
\end{tabular}
\\
$^\dag$Reipurth \& Aspin (2004): photometry is obtained in an aperture with radius 0.9\arcsec

$^\S$McGehee et al. (2004)

$^\ast$Ojha et al. (2005): IRSF photometry is obtained in an aperture with
radius 7.2\arcsec.

$^\pm$Upper limit in $K$-band is given (source is saturated in $K$-band).
\end{table}


\begin{table}
\caption{Spectroscopic observations of V1647 Ori at HCT}
\begin{tabular}{cccc}
\\
\hline
\hline
Date (UT)   &    JD     & Resolution (\AA) & Exposure (sec)\\
\hline
2004 Feb 22 & 2453057.9 &      7         & 900\\
2004 Feb 23 & 2453059.2 &      7         & 900\\
2004 Nov 16 & 2453326.5 &     15         & 900\\
2004 Nov 19 & 2453329.4 &     15         & 900\\
2005 Jan 8  & 2453379.2 &      7         & 900\\
2005 Jan 9  & 2453380.2 &      7         & 900\\
2005 Jan 16 & 2453387.1 &      7         & 900\\
2005 Mar 28 & 2453458.2 &      7         & 900\\
2005 Sep 8  & 2453621.7 &      7         & 4500\\
2005 Sep 28 & 2453642.5 &      7         & 1800\\
\hline
\end{tabular}
\end{table}


\begin{table}
\caption{$BVRI$ photometry of reference star A at HCT}
\begin{tabular}{cccccc}
\\
\hline
\hline
Reference star A  &   $B$ &  $V$          &     $R$       &      $I$      & Date (UT)\\
\hline
USNO-A2.0 0825-01669040  &  ...  & 15.69$\pm$.01 & 14.93$\pm$.02 & 14.24$\pm$.02 & 2004 Feb 14$^\dag$\\
                 & 16.89$\pm$.01 & 15.65$\pm$.01 & 14.88$\pm$.02 & 14.23$\pm$.01 & 2004 Feb 22$^\dag$\\   
                 & 16.93$\pm$.04 & 15.67$\pm$.03 & 14.88$\pm$.03 & 14.20$\pm$.03 & 2004 Oct 18$^\dag$\\
                 & 16.93$\pm$.02 & 15.63$\pm$.02 & 14.88$\pm$.01 & 14.23$\pm$.02 & 2004 Nov 17$^\dag$\\
                 & 16.92$\pm$.02 & 15.68$\pm$.01 & 14.86$\pm$.02 & 14.23$\pm$.01 & 2004 Dec 7$^\dag$\\
                 &     ...       & 15.65$\pm$.01 & 14.89$\pm$.01 &      ...      & 2005 Sep 26$^\dag$\\
                 & 16.94$\pm$.01 & 15.66$\pm$.01 & 14.89$\pm$.01 &...
& 2005 Sep 29$^\dag$\\
                 &     ...       & 15.67$\pm$.01 & 14.91$\pm$.01 &
14.23$\pm$.01    & 2005 Nov 28$^\dag$\\
                 &      ...      & 15.69$\pm$.01 & 14.92$\pm$.01 & 14.25$\pm$.01 &
2005 Nov 30$^\dag$\\  
\\
                 & 16.92$\pm$.01 & 15.66$\pm$.01 & 14.87$\pm$.01 & 14.21$\pm$.01 & 2004 Feb$^\S$\\
                 &  ...          & 15.65         & 14.91         & 14.26         & 2004 Apr 24$^\pm$\\
                 &  ...          & 15.66$\pm$.02 & 14.89$\pm$.02 & 14.23$\pm$.05 & 2004 Aug -- Oct$^\ast$\\
\hline
\end{tabular}

$^\dag$From HCT

$^\S$\mbox{From Henden's table (2004) of photometric measurements around McNeil's nebula}

$^\pm$from Kun et al. (2004)

$^\ast$from Semkov (2004)
\end{table}


\begin{table}
\caption{H$\alpha$ ($\lambda$6563), OI ($\lambda$7773) and Ca II IR triplet 
($\lambda$8498, $\lambda$8542 and $\lambda$8662) 
measurements from HCT spectra}
\begin{tabular}{cccccc}
\\
\hline
\hline
Date (UT) & W$_{\lambda}$ (6563) & W$_{\lambda}$ (OI 7773) & W$_{\lambda}$ (8498) & W$_{\lambda}$ (8542) & W$_{\lambda}$ (8662)\\
\hline
2004 Feb 22 & -34.59 & 3.82 & -8.67  & -9.56  &  -7.72\\
2004 Feb 23 & -33.62 & 2.63 & -7.67  & -7.84  &  -6.58\\
2004 Nov 16 & -27.25 & 5.41 & -12.91 & -10.14 & -12.39\\
2004 Nov 19 & -41.73 & 4.02 & -15.62 & -14.79 & -12.16\\
2005 Jan 8  & -28.81 & 2.81 & -9.70  & -9.08  & -7.20\\
2005 Jan 9  & -27.72 & 2.74 & -9.34  & -9.09  & -7.17\\
2005 Jan 16 & -32.54 & 2.69 & -9.66  & -10.17 &  -8.68\\
2005 Mar 28 & -20.34 & 2.98 & -10.66 &  -9.62 & -11.23\\
2005 Sep 8  & -26.16 & 2.58 & -16.73 & -19.10 & -18.43\\
$^\pm$2005 Sep 28 & -40.75 & ...  & -18.43 & -21.11 & -19.30\\
\hline
\end{tabular}

$^\pm$OI line is not clearly seen.
\end{table}  

\clearpage

\begin{figure}
\centering
\caption{$V H\alpha I$ three-colour composite image of McNeil's nebula ($V$: blue, 
$H\alpha$: green, $I$: red) obtained with HFOSC mounted on the 2m HCT telescope on 
2004 February 12. FOV is $\sim$ 10\arcmin x 10\arcmin. North is up and east is to 
the left. The location of V1647 Ori is marked at the centre of the image. The 
well known Herbig-Haro objects, such as the HH 24 complex, are seen in green colour 
because of the $H\alpha$ emission. The diffuse $H\alpha$ emission due to HH 23,
which may be driven by V1647 Ori, is marked by the box at top centre. The star
marked with A, is a reference star (see the text).  
\label{fig1}}
\end{figure}

\clearpage

\begin{figure}
\caption{(a) Enlarged view of $V H\alpha I$ three-colour composite image of  
McNeil's nebula (see figure 1). (b) $JHK$ three-colour composite image of McNeil's
nebula ($J$: blue, $H$: green, $K$: red) obtained with NIRCAM mounted on the 2m HCT 
telescope on 2004 October 18. North is up and east is to the left.
\label{fig2}}
\end{figure}

\begin{figure}
\caption{(a) Compact nebula seen around the illuminating star of McNeil's 
nebula in $K$-band image obtained on 2004 October 18. The lowest contour
and the contour interval are 18.0 mag arcsec$^{-2}$ and 0.5 mag arcsec$^{-2}$,
respectively. (b) PSF-subtracted image of McNeil's nebula region in the 
$K$-band displayed on a logarithmic scale. North is up and east is to the left.
The locations of the NIR central source (V1647 Ori) and the star $\sim$ 30\arcsec
~south-west of V1647 Ori, are indicated by filled triangles.
\label{fig3}}  
\end{figure}

\clearpage
\begin{figure}
\caption{NIR and optical images of the region surrounding McNeil's nebula. HCT
$K$-band image (grey colour) is overlaid with contours from the optical $H\alpha$ 
image obtained on 2004 October 18. North is up and east is to the left.
\label{fig4}}
\end{figure} 

\clearpage
\begin{figure}
\includegraphics[width=175mm]{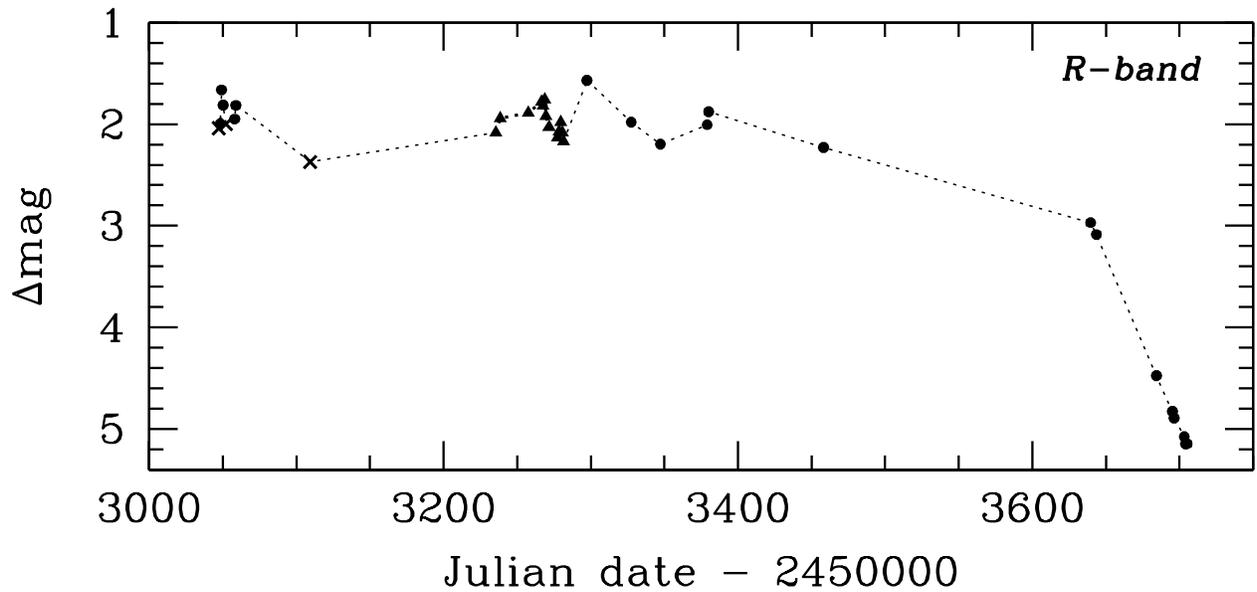}
\vspace*{-9.2cm}
\caption{The optical light curve of V1647 Ori in $R$-band. The differential magnitudes
are relative to a single comparison star (see the text), which appears 
constant to within 0.05 mag. The filled circles show our measurements 
(2004 February -- 2005 November). The cross symbols show the photometric 
measurements from McGehee et al. (2004). The filled triangles are the 
$R$-band photometric observations by Semkov (2004) during the 
period 2004 August -- October. 
\label{fig5}}
\end{figure}

\begin{figure}
\centering
\includegraphics[width=98mm]{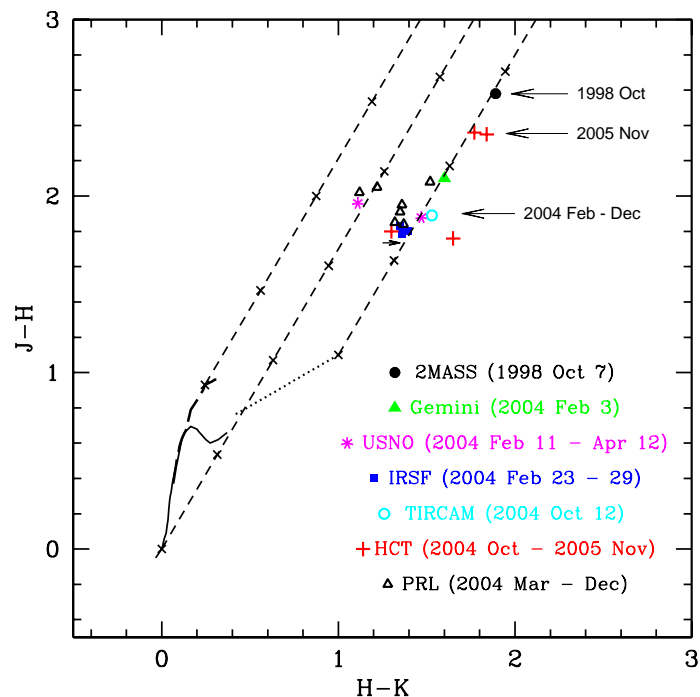}
\vspace*{-0.1cm}
\caption{$J-H/H-K$ colour-colour diagram showing the location of V1647 Ori
as observed with 2MASS (filled circle) on 1998 October 7; with IRSF (filled
squares) on 2004 February 23 -- 29; with Gemini (filled triangle) on 
2004 February 3; with USNO (asterisks) on 2004 February 11 and April 12; 
with TIRCAM (open circle) on 2004 October 12; with HCT (plus symbols) on 
2004 October -- 2005 November; 
and with NICMOS (open triangles) on 
2004 March -- December. The left small arrow shows the limit on $H-K$ colour on 
2004 November 19 (HCT) due to the saturation in $K$-band. The sequences of field 
dwarfs (solid curve) and giants (thick dashed curve) are from Bessell and 
Brett (1988). The dotted line represents the locus of T Tauri stars 
(Meyer et al. 1997). The dashed straight lines represent the reddening 
vectors (Rieke \& Lebofsky 1985). The crosses on the dashed lines are 
separated by $A_V$ = 5 mag.
\label{fig6}}
\end{figure}

\clearpage
\begin{figure}
\includegraphics[width=175mm]{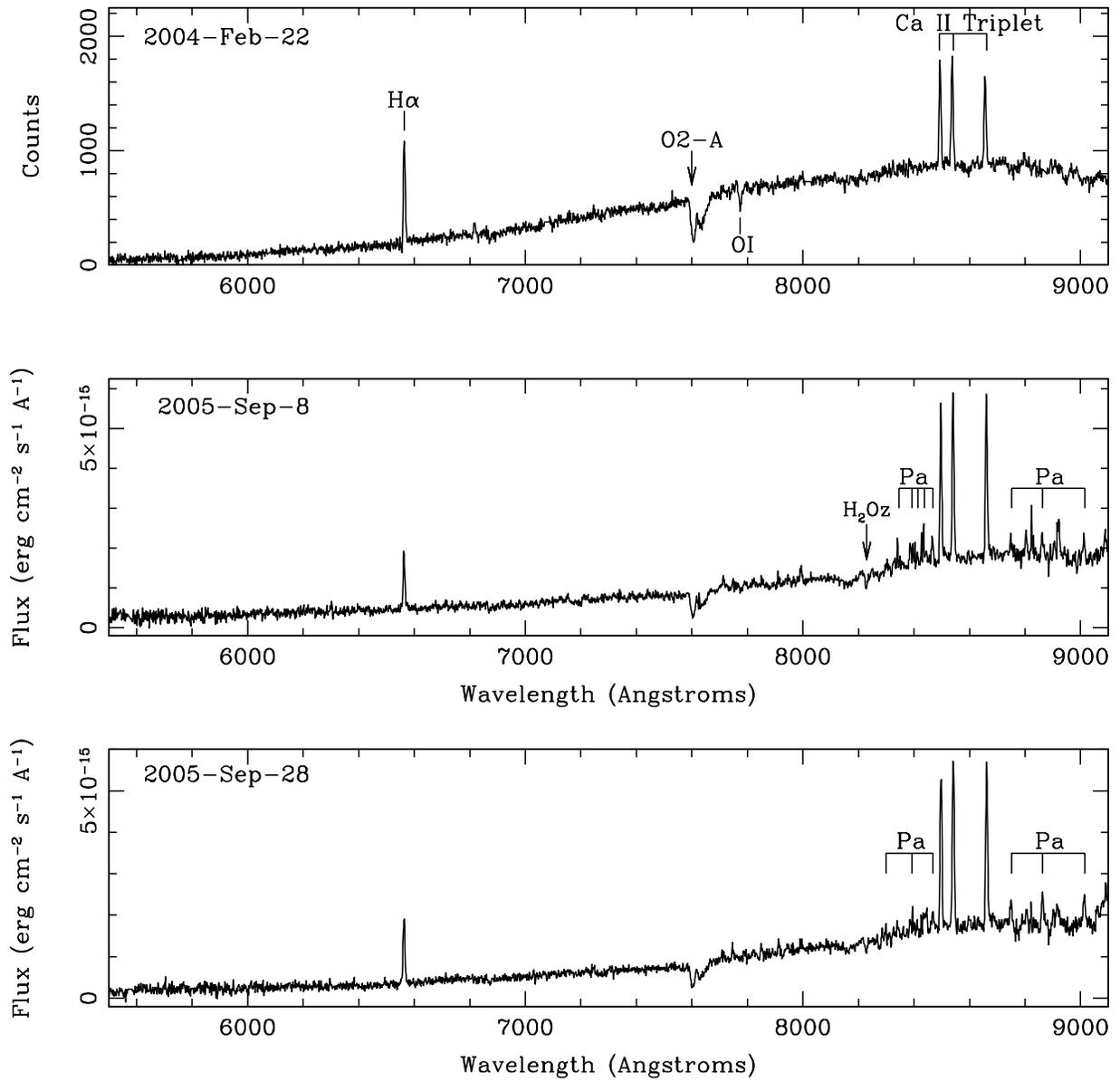}
\caption{Optical spectra of V1647 Ori obtained with HFOSC at the 2m HCT telescope
in the wavelength range 5500 - 9100\AA. 
\label{fig7}} 
\end{figure}

\clearpage
\begin{figure}
\includegraphics[width=175mm]{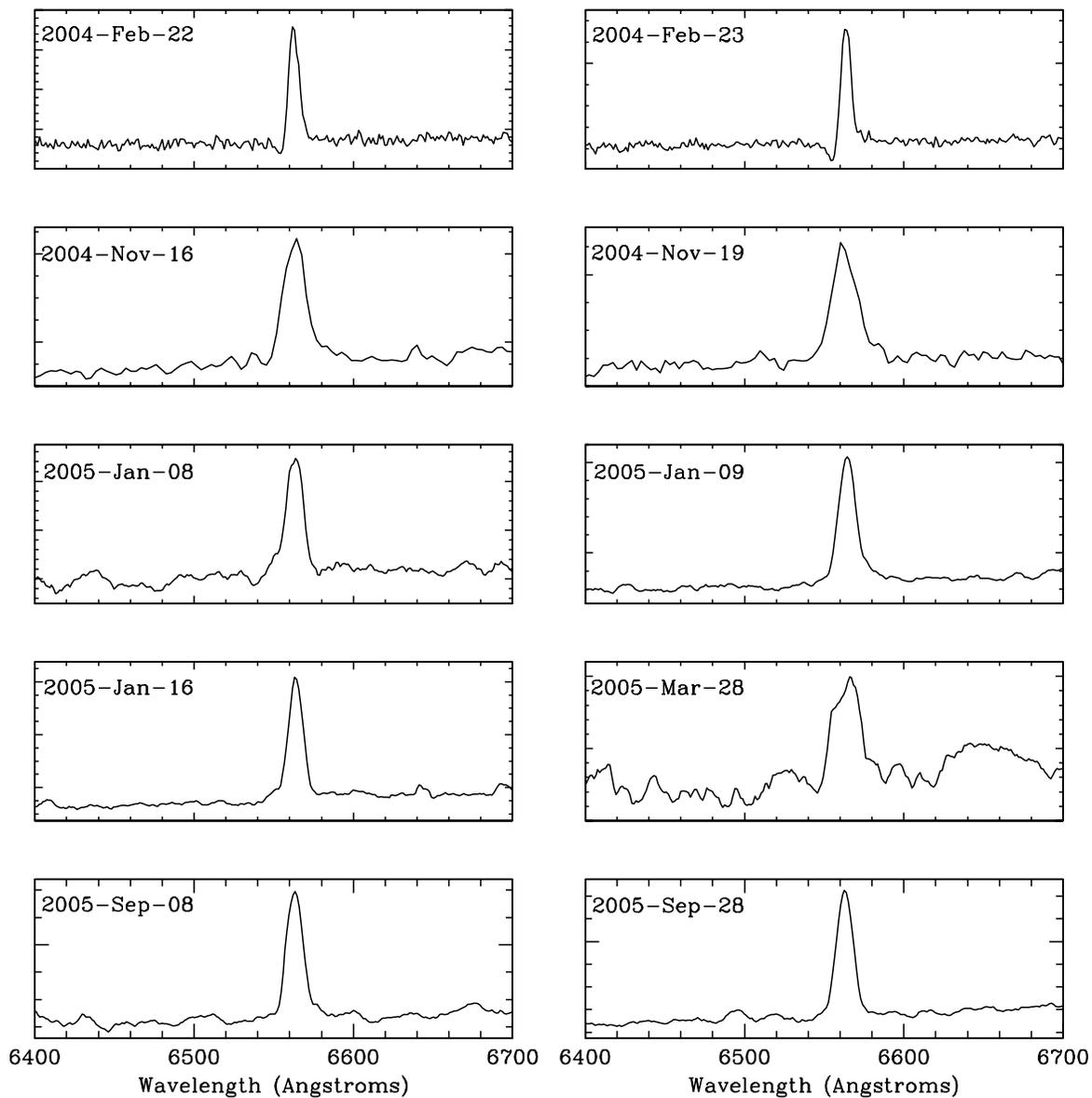}
\caption{The H$\alpha$ profiles of V1647 Ori seen in the spectra obtained with 
HFOSC between 2004 February 22 and 2005 September 28. The H$\alpha$ line shows
a pronounced blue-shifted P Cygni profile in the spectra taken in the early
phase of outburst (2004 February 22 and 23). 
\label{fig8}}
\end{figure}

\clearpage
\begin{figure}
\includegraphics[width=175mm]{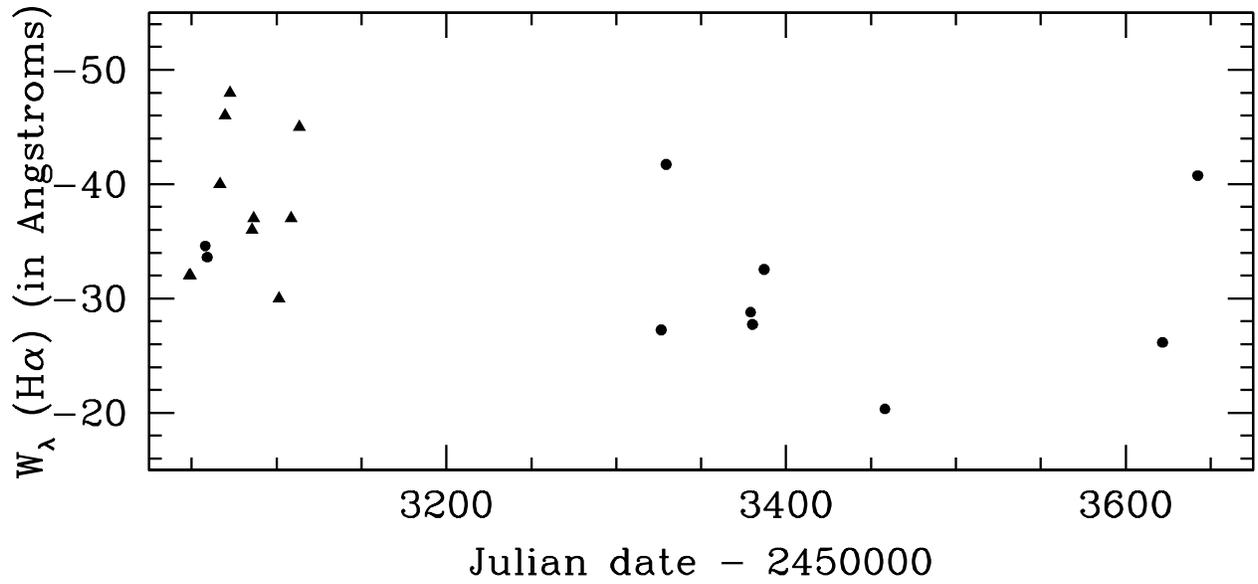}
\vspace*{-9cm}
\caption{The time variation of the $H\alpha$ equivalent width. The filled 
circles denote our measurements with HCT (2004 February 22 -- 2005 September 28). 
The filled
triangles show the data points (2004 February 13 -- April 17) from Walter et al 
(2004). While the data from Walter et al (2004) are suggestive of an oscillatory 
or periodic variation, we probably also see such a variation in our dataset.
\label{fig9}}   
\end{figure}

\end{document}